\documentstyle[prd,aps,amssymb,amsmath,amsthm]{revtex}

\newcommand{\half}{\mbox{$\textstyle \frac{1}{2}$}}
\newcommand{\quat}{\mbox{$\textstyle \frac{1}{4}$}}
\newcommand{\octa}{\mbox{$\textstyle \frac{1}{8}$}}

\tighten
\begin{document}
\draft \preprint{Imperial/TP/01-02/**}
\twocolumn[\hsize\textwidth\columnwidth\hsize\csname
@twocolumnfalse\endcsname

\title{Efficient Simulation of Quantum State Reduction}

\author{Dorje C. Brody
\footnote[1]{Electronic address: dorje@ic.ac.uk} and Lane P.
Hughston \footnote[2]{Electronic address: lane.hughston@kcl.ac.uk}
}

\address{${}^*$Blackett Laboratory, Imperial College,
London SW7 2BZ, UK}
\address{${}^\dagger$Department of Mathematics, King's College
London, The Strand, London WC2R 2LS, UK}

\date{\today}

\maketitle

\begin{abstract}
The energy-based stochastic extension of the Schr\"odinger
equation is a rather special nonlinear stochastic differential
equation on Hilbert space, involving a single free parameter, that
has been shown to be very useful for modelling the phenomenon of
quantum state reduction. Here we construct a general closed form
solution to this equation, for any given initial condition, in
terms of a random variable representing the terminal value of the
energy and an independent Brownian motion. The solution is
essentially algebraic in character, involving no integration, and
is thus suitable as a basis for efficient simulation studies of
state reduction in complex systems.
\end{abstract}


\vskip2pc]

The standard energy-based stochastic extension of the
Schr\"odinger equation is given by the following stochastic
differential equation:
\begin{eqnarray}
{\rm d}|\psi_t\rangle &=& -{\rm i}{\hat H} |\psi_t\rangle {\rm d}t
- \octa \sigma^2({\hat H}-H_t)^2 |\psi_t\rangle {\rm d}t \nonumber
\\ & & + \half \sigma ({\hat H}-H_t)|\psi_t\rangle {\rm d}W_t ,
\label{eq:1.1}
\end{eqnarray}
with initial condition $|\psi_0\rangle$. Here $|\psi_t\rangle$ is
the state vector at time $t$, ${\hat H}$ is the Hamiltonian
operator, $W_t$ denotes a one-dimensional Brownian motion, and
\begin{eqnarray}
H_t = \frac{\langle{\psi}_t|{\hat H}|\psi_t\rangle}
{\langle{\psi}_t|\psi_t\rangle}  \label{eq:1.2}
\end{eqnarray}
is the expectation of ${\hat H}$ in the state $|\psi_t\rangle$.
The parameter $\sigma$, which has the units $ \sigma\sim[{\rm
energy}]^{-1}[{\rm time}]^{-1/2}$, governs the characteristic
timescale $\tau_R$ associated with the collapse of the wave
function induced by (\ref{eq:1.1}). This is given by $ \tau_R =
1/\sigma^2 V_0$, where $V_0$ is the initial value of the squared
energy uncertainty, which at time $t$ is
\begin{eqnarray}
V_t = \frac{\langle{\psi}_t|({\hat H}-H_t)^2|\psi_t\rangle}
{\langle{\psi}_t|\psi_t\rangle}.  \label{eq:1.5}
\end{eqnarray}

The stochastic equation (\ref{eq:1.1}) provides perhaps the
simplest known physically plausible model for state vector
reduction in quantum mechanics \cite{gisin,hughston}. Although its
properties have been studied extensively, it has hitherto been
necessary to resort to numerical methods to solve (\ref{eq:1.1}).
The purpose of this article is to present an analytic solution for
the dynamics of $|\psi_t\rangle$. Apart from its use as a means
for generating a general solution to a nonlinear problem in
quantum state dynamics, the method we propose also sheds new light
on the nature of quantum probability and some of the issues
associated with the flow of information when quantum measurements
are made.

We begin with a brief overview of the stochastic framework
implicit in the extended Schr\"odinger dynamics given by equation
(\ref{eq:1.1}). We follow closely here the analysis presented in
\cite{brody1}. Specifically, we introduce first the key notions of
filtration, conditional expectation, martingale, and potential. We
then demonstrate that the conditional expectation (\ref{eq:4.2})
gives rise to the energy expectation process (\ref{eq:1.2}). As a
consequence, we are led to simple analytic expressions for the
energy (\ref{eq:4.14}) and the state vector (\ref{eq:26}) in terms
of a pair of underlying state variables. These results open up the
possibility of efficiently simulating the reduction process for a
variety of models. Finally we illustrate the practical advantages
of our method by analysing in some detail the timescale associated
with the reduction process in the case of a two-state system.

The dynamics of $|\psi_t\rangle$ are defined on a probability
space $(\Omega,{\mathcal F},{\mathbb P})$ with filtration
${\mathcal F}_t$ ($0\leq t<\infty$). Here $\Omega$ is the sample
space over which ${\mathcal F}$ is a $\sigma$-field of open sets
upon which the probability measure ${\mathbb P}$ is defined.

The filtration represents the information available at time $t$.
More specifically, a filtration of ${\mathcal F}$ is a collection
$\{{\mathcal F}_t\}$ of $\sigma$-subfields of ${\mathcal F}$ such
that $s\leq t$ implies ${\mathcal F}_s\subset {\mathcal F}_t$.
Given a random variable $X$ on $(\Omega,{\mathcal F},{\mathbb P})$
for which ${\mathbb E}[X]$ exists, we write ${\mathbb
E}[X|{\mathcal F}_t]$ for the {\sl conditional expectation} of $X$
with respect to the $\sigma$-subfield ${\mathcal
F}_t\subset{\mathcal F}$. Intuitively, conditioning with respect
to ${\mathcal F}_t$ means giving the information available up to
time $t$. The nesting ${\mathcal F}_s\subset{\mathcal F}_t$ for
$s\leq t$ thus gives rise to a notion of causality. For
convenience, we use the abbreviation ${\mathbb E}_t[X]={\mathbb
E}[X|{\mathcal F}_t]$ when the choice of $\{{\mathcal F}_t\}$ is
understood. The conditional expectation ${\mathbb E}_t[X]$
satisfies: (i) the tower property ${\mathbb E}_s[{\mathbb E}_t
[X]]={\mathbb E}_s[X]$ for $s\leq t$; and (ii) the law of total
probability ${\mathbb E}[{\mathbb E}_t[X]] ={\mathbb E}[X]$. If
${\mathbb E}_t[X]=X$ we say that $X$ is ${\mathcal F}_t$-{\sl
measurable}.

The conditional expectation operation allows us to introduce the
concept of a martingale, the stochastic analogue of a conserved
quantity. A process $X_t$ is said to be an ${\mathcal F}_t$-{\em
martingale} if ${\mathbb E}[|X_t|]<\infty$ and ${\mathbb
E}_s[X_t]=X_s$ for all $0\leq s\leq t<\infty$. In other words,
$X_t$ is an ${\mathcal F}_t$-martingale if it is integrable and if
its conditional expectation, given information up to time $s$, is
the value $X_s$ of the process at that time. If the filtration is
fixed, then we can simply speak of a martingale without further
qualification. There are circumstances, however, where more than
one filtration can enter a problem, and then we have to specify
with respect to which filtration the martingale property holds.

For a concise mathematical representation of the state reduction
process, we also require the concepts of supermartingale and
potential. A process $X_t$ is an ${\mathcal F}_t$-{\em
supermartingale} if ${\mathbb E}[|X_t|]<\infty$ and ${\mathbb
E}_s[X_t]\leq X_s$ for all $0\leq s\leq t<\infty$. Intuitively, a
supermartingale is on average a nonincreasing process. A positive
supermartingale $X_t$ with the property ${\mathbb E}[X_t]
\rightarrow0$ as $t\rightarrow\infty$ is called a {\sl potential}.

The filtration ${\mathcal F}_t$ with respect to which stochastic
extension of the Schr\"odinger equation (\ref{eq:1.1}) is defined
is generated in a standard way by the Wiener process $W_t$. We
signify this by writing ${\mathcal F}_t={\mathcal F}_t^W$. It is
straightforward to verify that the Hamiltonian process $H_t$ is an
${\mathcal F}_t^W$-martingale, and that the variance process $V_t$
is an ${\mathcal F}_t^W$-supermartingale. That is to say,
${\mathbb E}_s[H_t]=H_s$, and ${\mathbb E}_s[V_t] \leq V_s$. These
relations can be deduced by applying Ito's lemma to (\ref{eq:1.2})
and (\ref{eq:1.5}), from which we infer that
\begin{eqnarray}
H_t = H_0 + \sigma \int_0^t V_s {\rm d}W_s \label{eq:2.5}
\end{eqnarray}
and
\begin{eqnarray}
V_t = V_0 -\sigma^2 \int_0^t V_s^2 {\rm d}s + \sigma \int_0^t
\beta_s {\rm d}W_s . \label{eq:2.6}
\end{eqnarray}
Here $ \beta_t = \langle{\psi}_t|({\hat H}-H_t)^3
|\psi_t\rangle/\langle{\psi}_t|\psi_t\rangle$ is the {\sl
skewness} of the energy distribution at time $t$. The martingale
and the supermartingale relations then follow as a consequence of
elementary properties of the stochastic integrals appearing in
(\ref{eq:2.5}) and (\ref{eq:2.6}).

In the case of the ordinary Schr\"odinger equation with a
time-independent Hamiltonian, the energy process (\ref{eq:1.2}) is
constant. This is usually interpreted as the quantum mechanical
expression of an energy conservation law. However, if a system is
in an indefinite state of energy then it is not clear {\it a
priori} what is meant by energy conservation. The martingale
condition ${\mathbb E}_s[H_t]=H_s$ can be interpreted as a
generalised energy conservation law applicable in such
circumstances. In particular, it implies that once the state
reduction has occurred, the probabilistic average of the outcome
for the energy must equal the initial expectation.

The supermartingale property satisfied by $V_t$ on the other hand
is the essence of what is meant by a {\sl reduction} process. In
fact, it follows from equation (\ref{eq:2.6}) that the asymptotic
behaviour of $V_t$ is given by $\lim_{t\rightarrow\infty}{\mathbb
E}\left[V_t\right] = 0$. In other words, the variance process for
the energy is a potential. Writing $ H_\infty = H_0 + \sigma
\int_0^\infty V_t {\rm d}W_t$ for the random terminal value of the
energy, one can prove \cite{brody1} as a consequence of
(\ref{eq:2.5}) and (\ref{eq:2.6}) that
\begin{eqnarray}
H_t = {\mathbb E}_t[H_\infty] \label{eq:2.8}
\end{eqnarray}
and that
\begin{eqnarray}
V_t = {\mathbb E}_t[(H_\infty-H_t)^2] . \label{eq:2.10}
\end{eqnarray}
That is to say, $H_t$ and $V_t$ are respectively the ${\mathcal
F}_t^W$-conditional mean and variance of $H_\infty$.

With these facts in hand, we now present a method for obtaining a
general solution to the stochastic equation (\ref{eq:1.1}). The
setup is as follows. We denote by $E_i$ ($i=1,2,\ldots$) the
eigenvalues of the Hamiltonian of a given quantum system, and
write
\begin{eqnarray}
\pi_i =  \frac{|\langle\psi_0|\psi_i\rangle|^2} {\langle\psi_0
|\psi_0\rangle\langle\psi_i|\psi_i\rangle} , \label{eq:3.15}
\end{eqnarray}
for the transition probability from the given initial state
$|\psi_0\rangle$ to the eigenstate $|\psi_i\rangle$ with energy
$E_i$. If the spectrum of ${\hat H}$ is degenerate, then
$|\psi_i\rangle$ denotes the L\"uders state, i.e. the projection
of $|\psi_0\rangle$ onto the linear subspace of states
corresponding to the eigenvalue $E_i$.

Now let the probability space $(\Omega,{\mathcal F},{\mathbb P})$
be given, and on it specify a random variable $H$ that takes the
values $E_i$ with probabilities $\pi_i$. We also assume that
$(\Omega,{\mathcal F},{\mathbb P})$ comes equipped with a
filtration ${\mathcal G}_t$ with respect to which a standard
Brownian motion $B_t$ is specified, and that $H$ and $B_t$ are
{\sl independent}. We assign no {\it a priori} physical
significance to $H$ and $B_t$, which are introduced as an ansatz
for obtaining a solution for (\ref{eq:1.1}).

We now define a random process $\xi_t$, which we shall call the
signal process, according to the scheme
\begin{eqnarray}
\xi_t = \sigma H t + B_t . \label{eq:4.1}
\end{eqnarray}
Intuitively, one can think of $\xi_t$ as giving a `noisy'
representation of the information encoded in the random variable
$H$.

We let $\{{\mathcal F}_t^\xi\}$ denote the filtration generated by
the process $\xi_t$, i.e. the information generated by $\xi_t$ as
time progresses, and consider the conditional expectation
\begin{eqnarray}
H_t = {\mathbb E}\left[ H|{\mathcal F}_t^\xi\right] .
\label{eq:4.2}
\end{eqnarray}
Clearly, ${\mathcal F}_t^\xi\subset{\mathcal G}_t$ since knowledge
of $H$ together with $\{B_s\}_{0\leq s\leq t}$ implies knowledge
of $\{\xi_s\}_{0\leq s\leq t}$, although the converse is not the
case.

The significance of the ${\mathcal F}_t^\xi$-martingale $H_t$ is
that it represents the `best estimate' for the value of $H$ given
the history of the signal process $\xi_s$ from time $0$ up to time
$t$. More precisely, an ${\mathcal F}_t^\xi$-measurable random
variable $Y_t$ minimises the expectation of the squared deviation
of $H$ from $Y_t$, given ${\mathcal F}_t^\xi$, iff $Y_t = {\mathbb
E}[H|{\mathcal F}_t^\xi]$. This follows, by a variational
argument, from the relation ${\mathbb E}[(H-Y_t)^2|{\mathcal
F}_t^\xi] = {\mathbb E}[H^2|{\mathcal F}_t^\xi] -2 Y_t {\mathbb
E}[H|{\mathcal F}_t^\xi] + Y_t^2$.

We proceed to establish the remarkable fact that {\sl the process
$H_t$ defined by (\ref{eq:4.2}) is statistically indistinguishable
from the energy process (\ref{eq:1.2}) associated with the
stochastic extension of the Schr\"odinger equation
(\ref{eq:1.1})}.

The argument goes as follows. First, because $\xi_t$ is a Markov
process satisfying $\lim_{t\rightarrow\infty} t^{-1}\xi_t=H$, we
have
\begin{eqnarray}
{\mathbb E}\left[ H|{\mathcal F}_t^\xi\right] = {\mathbb E}\left[
H|\xi_t\right]. \label{eq:4.2-2}
\end{eqnarray}
In other words, to determine the conditional expectation of $H$
given $\{\xi_s\}_{0\leq s\leq t}$ it suffices to condition on
$\xi_t$ alone.

To calculate ${\mathbb E}[H|\xi_t]$, we require a version of the
Bayes formula applicable when we consider the probability of a
discrete random variable conditioned on the value of a continuous
random variable. In particular,
\begin{eqnarray}
{\mathbb P}(H=E_i|\xi_t) &=& \frac{\pi_i\rho (\xi_t|H=E_i)}
{\sum_{i} \pi_i \rho(\xi_t|H=E_i)}, \label{eq:4.12}
\end{eqnarray}
where $\pi_i={\mathbb P}(H=E_i)$. Here $\rho(\xi_t|H=E_i)$ denotes
the conditional probability density for the continuous random
variable $\xi_t$ given that $H=E_i$. Since $B_t$ is a standard
Brownian motion, the conditional density for $\xi_t$ is
\begin{eqnarray}
\rho(\xi_t|H=E_i) = \frac{1}{\sqrt{2\pi t}} \exp\left(
\frac{1}{2t} (\xi_t-\sigma E_i t)^2\right) . \label{eq:4.13}
\end{eqnarray}
It follows from the Bayes law (\ref{eq:4.12}) that the
conditional probability for the random variable $H$ is
\begin{eqnarray}
{\mathbb P}(H=E_i|\xi_t) = \frac{\pi_i \exp\left( \sigma E_i \xi_t
-\half \sigma^2 E_i^2 t\right)}{\sum_{i} \pi_i \exp\left( \sigma
E_i \xi_t-\half \sigma^2 E_i^2 t\right)}. \label{eq:4.133}
\end{eqnarray}
Therefore, we deduce that the conditional expectation of $H$ given
$\xi_t$ is
\begin{eqnarray}
H_t &=& \sum_{i} E_i {\mathbb P}(H=E_i|\xi_t) \nonumber \\
&=&  \frac{\sum_{i} \pi_i E_i \exp\left( \sigma E_i \xi_t-\half
\sigma^2 E_i^2 t\right)}{\sum_{i} \pi_i \exp\left( \sigma E_i
\xi_t-\half \sigma^2 E_i^2 t\right)}. \label{eq:4.14}
\end{eqnarray}

In order to show that $H_t$ is the energy process of the given
quantum system, one further key result is required: namely, that
{\sl the process $W_t$ defined by
\begin{eqnarray}
W_t = \xi_t - \sigma \int_0^t H_s {\rm d}s \label{eq:4.17}
\end{eqnarray}
is an ${\mathcal F}_t^\xi$-Brownian motion}. To verify this, it
suffices, by virtue of L\'evy's characterisation of Brownian
motion \cite{shiryaev}, to demonstrate ($a$) that $W_t$ is an
${\mathcal F}_t^\xi$-martingale, and ($b$) that $({\rm
d}W_t)^2={\rm d}t$. To verify ($b$) we note that (\ref{eq:4.1})
implies ${\rm d} \xi_t =\sigma H{\rm d}t+{\rm d} B_t$, and thus
$({\rm d}\xi_t)^2={\rm d}t$. On the other hand, (\ref{eq:4.17})
implies that ${\rm d}W_t={\rm d}\xi_t- \sigma H_t{\rm d}t$, and
hence $({\rm d}W_t)^2=({\rm d}\xi_t)^2$. To establish ($a$), let
(\ref{eq:4.14}) define a function $H(\xi,t)$ of two variables such
that $H_t=H(\xi_t,t)$:
\begin{eqnarray}
H(\xi,t) = \frac{\sum_{i} \pi_i E_i \exp\left( \sigma E_i
\xi-\half \sigma^2 E_i^2 t\right)}{\sum_{i} \pi_i \exp\left(
\sigma E_i \xi-\half \sigma^2 E_i^2 t\right)}. \label{eq:4.144}
\end{eqnarray}
Then applying Ito's lemma and using the relation $({\rm
d}\xi_t)^2={\rm d}t$, we obtain
\begin{eqnarray}
{\rm d}H_t = \left( \partial_t H(\xi_t,t)+ \half \partial_\xi^2
H(\xi_t,t)\right) {\rm d}t + \partial_\xi H(\xi_t,t){\rm d}\xi_t
\label{eq:4.188}
\end{eqnarray}
where $ \partial_t H(\xi_t,t)$ denotes $ \partial H(\xi,t)/
\partial t$ valued at $\xi=\xi_t$, and so on. A short
calculation making use of (\ref{eq:4.144}) shows that
$\partial_\xi H(\xi_t,t) = \sigma V(\xi_t, t)$ and $\partial_t
H(\xi_t,t) + \half \partial_\xi^2 H(\xi_t,t) = -\sigma^2
V(\xi_t,t) H(\xi_t,t)$, where the function $V(\xi,t)$ is
\begin{eqnarray}
V(\xi,t) &=& \frac{\sum_{i} \pi_i (E_i-H(\xi,t))^2 \exp\left(
\sigma E_i \xi-\half \sigma^2 E_i^2 t\right)}{\sum_{i} \pi_i
\exp\left( \sigma E_i \xi-\half \sigma^2 E_i^2 t\right)}.
\label{eq:4.22}
\end{eqnarray}
It follows, in particular, that $H(\xi,t)$ is monotonic in $\xi$
for any given value of $t$. Substituting these results into
(\ref{eq:4.188}), we infer that ${\rm d}H_t = \sigma V(\xi_t,t)
{\rm d}W_t$. However, we know that $H_t$ is an ${\mathcal
F}_t^\xi$-martingale, and therefore we conclude that $W_t$ is also
an ${\mathcal F}_t^\xi$-martingale, and that establishes ($a$). We
thus deduce that $W_t$ is an ${\mathcal F}_t^\xi$-Brownian motion,
as claimed. We call $W_t$ the {\sl innovation process} associated
with $H_t$. The significance of $W_t$ is that the process $\xi_t$
defined in (\ref{eq:4.1}) satisfies the diffusion equation ${\rm
d}\xi_t = \sigma H_t {\rm d}t + {\rm d}W_t$, where
$H_t=H(\xi_t,t)$.

Now let $|\psi_0\rangle$ be the initial normalised state vector of
the quantum system, and let ${\hat P}_i$ denote for each value of
$i$ the projection operator onto the Hilbert subspace
corresponding to the energy eigenvalue $E_i$. We let
$|\psi_i\rangle=\pi_i^{-1/2}{\hat P}_i |\psi_0\rangle$ denote the
L\"uders state corresponding to $E_i$, and write $\Pi_{it} =
{\mathbb P}\left( H=E_i|\xi_t\right)$ for the process defined by
(\ref{eq:4.133}). Then, we can verify that
\begin{eqnarray}
|\psi_t\rangle = \sum_{i} {\rm e}^{-{\rm i}E_i t} \Pi_{it}^{1/2}
|\psi_i\rangle  \label{eq:4.24}
\end{eqnarray}
satisfies the stochastic extension of the Schr\"odinger equation
(\ref{eq:1.1}) with the given initial condition. In particular, by
applying Ito's lemma to (\ref{eq:4.133}) and using the diffusion
equation satisfied by $\xi_t$ we obtain
\begin{eqnarray}
{\rm d}\Pi_{it} = \sigma (E_i-H_t)\Pi_{it} {\rm d}W_t.
\label{eq:4.25}
\end{eqnarray}
With another application of Ito's lemma we deduce that $ {\rm
d}\Pi_{it}^{1/2} = - \octa \sigma^2 (E_i-H_t)^2 \Pi_{it}^{1/2}
{\rm d}t + \half \sigma (E_i-H_t)\Pi_{it}^{1/2} {\rm d} W_t$. A
short calculation then shows that (\ref{eq:4.24}) satisfies
(\ref{eq:1.1}), and that the expectation of the operator ${\hat
H}$ in the state $|\psi_t\rangle$ is the process (\ref{eq:4.14}).

Summing up, the stochastic equation (\ref{eq:1.1}) can be solved
as follows. We let $H$ be a random variable taking values $E_i$
with the probabilities $\pi_i$ defined by (\ref{eq:3.15}), or
equivalently $\pi_i^{1/2} = \langle\psi_0|{\hat P}_i
|\psi_0\rangle/\langle\psi_0|\psi_0\rangle$. Letting $B_t$ denote
an independent Brownian motion, we set $\xi_t=\sigma Ht+B_t$. The
solution of (\ref{eq:1.1}) is then given by
\begin{eqnarray}
|\psi_t\rangle = \frac{\sum_{i} \sqrt{\pi_i} \exp\left( -{\rm
i}E_i t+\frac{1}{2}\sigma E_i\xi_t - \frac{1}{4}\sigma^2 E_i^2
t\right)|\psi_i\rangle} {\left( \sum_{i} \pi_i \exp\left( \sigma
E_i\xi_t - \frac{1}{2}\sigma^2 E_i^2 t\right)\right)^{1/2}}
\label{eq:26}
\end{eqnarray}
where the ${\mathcal F}_t^\xi$-Brownian motion $W_t$ driving
$|\psi_t\rangle$ in (\ref{eq:1.1}) is given by (\ref{eq:4.17}). In
particular, by use of (\ref{eq:26}), the expression
(\ref{eq:4.14}) for $H_t$ follows at once since $\langle\psi_i|
{\hat H}|\psi_j\rangle=E_i \delta_{ij}$.

The fact that (\ref{eq:4.14}) is indeed a reduction process for
the energy can be verified directly as follows. Suppose, in a
particular realisation of the process $H_t$, the random variable
$H$ takes the value $E_j$ for some choice of the index $j$.
Setting $\omega_{ij}=E_i-E_j$ and writing $\xi_t=\sigma E_j
t+B_t$, we have, for the corresponding realisation of $H_t$,
\begin{eqnarray}
H_t &=& \frac{\sum_{i} \pi_i E_i \exp\left( \sigma E_i B_t-\half
\sigma^2 E_i(E_i-2E_j) t\right)}{\sum_{i} \pi_i \exp\left(
\sigma E_i B_t-\half \sigma^2 E_i\omega_{ij} t\right)} \nonumber \\
&=& \frac{\pi_j E_j + \sum_{i}' \pi_i E_i \exp\left( \sigma
\omega_{ij} B_t-\half \sigma^2 \omega_{ij}^2 t\right)}{\pi_j +
\sum_{i}' \pi_i \exp\left( \sigma \omega_{ij} B_t-\half \sigma^2
\omega_{ij}^2 t\right)}, \label{eq:4.27}
\end{eqnarray}
where $\sum_i'=\sum_{i(\neq j)}$. However, the exponential
martingale $M_{ijt}$ defined for $i\neq j$ by
\begin{eqnarray}
M_{ijt} = \exp\left( \sigma \omega_{ij}B_t - \half \sigma^2
\omega_{ij}^2 t\right) \label{eq:4.28}
\end{eqnarray}
that appears in expression (\ref{eq:4.27}) has the property:
$\lim_{t\rightarrow\infty} {\mathbb P}\left( M_{ijt}>0\right) =
0$. Hence from
\begin{eqnarray}
H_t = \frac{\pi_j E_j + \sum_{i}' \pi_i E_i M_{ijt}}
{\pi_j+\sum_{i}' \pi_i M_{ijt}}, \label{eq:4.30}
\end{eqnarray}
we see that $H_t$ converges to the value $E_j$ with probability
one. A similar argument allows us to verify that if $H=E_j$ then
for each value of $i$ we have $ \lim_{t\rightarrow\infty}
\Pi_{it}={\bf 1}_{\{i=j\}}$, where ${\bf 1}$ denotes the indicator
function, which shows that $|\psi_t\rangle$ converges to the
L\"uders state corresponding to the energy eigenvalue $j$ with
probability one \cite{brody1}.

Therefore, we see that the random variable $H$ can be identified
with the terminal value $H_\infty$ of the energy process. The fact
that $H$ is not ${\mathcal F}_t^W$-measurable for $t<\infty$
indicates that the `true value' of $H$ is `hidden' until the
reduction process is complete. On a related interpretational point
we note that in stochastic models for quantum state reduction it
is sometimes assumed that the driving process $W_t$ is in some way
`external' to the quantum system, representing, e.g., a noisy
environmental coupling. This assumption, however, is unnecessary:
as far as the flow of information is concerned, we have
\begin{eqnarray}
{\mathcal F}_t^W = {\mathcal F}_t^\xi = {\mathcal F}_t^H =
{\mathcal F}_t^{|\psi\rangle} ,
\end{eqnarray}
and it is thus perfectly consistent to regard the innovation
process $W_t$ as being endogenous.

The advantage of the expressions (\ref{eq:4.14}) and (\ref{eq:26})
is that $H_t$ and $|\psi_t\rangle$ are expressed {\sl
algebraically} in terms of the underlying random variable $H$ and
the independent Brownian motion $B_t$. These can be thought of as
representing independent {\sl state variables} for the reduction
dynamics. As a consequence, we are able to investigate properties
of the reduction process (\ref{eq:1.1}) directly without having to
resort to numerical integration. In particular, by use of
(\ref{eq:26}) a numerical simulation of the state reduction of
rather complex quantum systems is feasible, including cases for
which the Hamiltonian has a nondiscrete spectrum.

In conclusion let us analyse now in detail the timescale
associated with the reduction process. For simplicity, we consider
a two-state system with energy levels $E_1$ and $E_2$. The initial
state is given by $|\psi_0\rangle$, and the transition
probabilities to the energy eigenstates $|E_1\rangle$ and
$|E_2\rangle$ are given by $\pi_1$ and $\pi_2$.

Suppose a measurement of the energy is made, and we condition on
the outcome of the measurement being $E_1$. In that case,
according to (\ref{eq:4.30}), we have
\begin{eqnarray}
H_t = \frac{\pi_1 E_1 + \pi_2 E_2 M_{21t}} {\pi_1 + \pi_2
M_{21t}}, \label{eq:5.1}
\end{eqnarray}
where $M_{21t} = \exp\left( \sigma\omega_{21}B_t - \half \sigma^2
\omega_{21}^2 t \right)$. Writing $\beta = \quat \sigma^2
\omega_{21}^2$ for the parameter that determines the
characteristic rate of reduction, we can work out the probability
that $M_{21t}<{\rm e}^{-n}$ for some value of $n$. Since $B_t$ is
normally distributed with zero mean and variance $t$, we find that
\begin{eqnarray}
{\mathbb P}\left( M_{21t}<{\rm e}^{-n}\right) &=& {\mathbb P}
\left( B_t < \beta^{1/2} t - \half \beta^{-1/2}n \right) \nonumber
\\ &=& N\left( (\beta t)^{1/2} - \half n (\beta t)^{-1/2} \right)
\label{eq:5.4}
\end{eqnarray}
where $N(x)$ is the standard normal distribution function.

Therefore, for example, we see that provided $t>5\tau_{R}$, we
have ${\mathbb P}\left( M_{21t}<{\rm e}^{-10}\right) > \half$,
where $\tau_R=1/\beta$. In particular, as $H_t$ draws near $E_1$
we have the relation
\begin{eqnarray}
H_t-E_1 \sim \frac{\pi_2}{\pi_1} (E_2-E_1) M_{21t} .
\end{eqnarray}
Thus, after only a relatively few multiples of the characteristic
reduction timescale, the amount by which $H_t$ differs from $E_1$
will typically be reduced to a tiny fraction of the energy
difference $E_2-E_1$.

DCB acknowledges support from The Royal Society. LPH acknowledges
the Institute for Advanced Study, Princeton, for hospitality while
part of this work was carried out.

\end{document}